\documentclass[pdflatex,sn-basic]{sn-jnl}
\usepackage{graphicx}
\usepackage{multirow}
\usepackage{amsmath,amssymb,amsfonts}
\usepackage{amsthm}
\usepackage{xcolor}
\usepackage{textcomp}
\usepackage{manyfoot}
\usepackage{geometry}
\usepackage{float}
\usepackage{longtable}
\usepackage{array}
\usepackage{booktabs}
\usepackage{url}
\usepackage{xurl}
\begin{document}
\title[From Tokens to Credits to Outcomes]{From Tokens to Credits to Outcomes: Pricing AI Across Models, Applications, and Cloud Platforms}
\author*{\fnm{Pratik} \sur{Gupta}}\email{pratikgupta1088@gmail.com}
\affil{\orgname{Stripe}}
\abstract{Between 2023 and 2026, AI changed how software is priced. Traditional
SaaS charged per seat because serving one more user cost almost nothing.
Model inference has non-zero marginal cost, and usage varies by orders
of magnitude across customers, producing prepaid credits, metered usage,
hybrid seat-plus-usage models, and outcome pricing that supplement, and
in rare cases replace, seat-based pricing. These pricing models have not
been systematically characterized. This paper presents a source-traceable
study of public pricing artifacts - pricing pages and documentation -
across three cohorts: the full Forbes AI 50 (2026) as the AI-native core
sample, cloud platforms selling AI services included by a documented
rule, and a frozen purposive cohort of 9 distinctive designs. It
contributes (1) a dataset coding each offering's pricing primitive and
credit mechanics with dated primary sources, prepared for public release
with the paper, (2) a taxonomy of the primitives in use and their
distribution across cohorts, and (3) evidence-informed design requirements
for automated, configurable billing platforms, including ledger state
for credits, attribution pipelines for outcome pricing, and
configuration support for high pricing-change velocity. Subscription
with usage quota is the dominant observable model among AI-native
companies (55\% of observable AI-native rows; 45--61\% under sensitivity
analysis of hybrid boundary cases, 41--65\% when product-line coding is also varied). 23 of the 52 AI-native rows publish
no pricing. Hybrid products are better understood as compositions of
packaging anchor, allowance mechanism, and billable unit than as
mutually exclusive categories. The dataset is planned for release with
the paper; my employer is excluded from the sample.}
\begingroup
\renewcommand\thefootnote{}%
\footnotetext{Independent research conducted in a personal capacity; the views expressed are my own and do not represent Stripe.}
\endgroup
\maketitle

\hypertarget{introduction}{%
\section{Introduction}\label{introduction}}

Classic SaaS pricing assumed near-zero marginal cost: once built,
serving one more user cost little, so charging per seat approximated
value without metering anything. AI products broke that assumption.
Model inference generally incurs non-zero compute and infrastructure
costs, and usage varies by orders of magnitude across customers.
Cost-linked pricing itself is not new: payment processors have priced
card transactions as underlying cost plus a markup (interchange++ /
Cost+) for well over a decade. What is particularly salient in the
current setting is the migration of that established economic model into
software pricing, together with the engineering complexity of metering
token usage at scale - input and output rates, cached tokens, tool
calls, multimodal units, and upstream model prices that change under the
bill. A uniform seat price can cause high-usage customers to consume
substantially more resources than low-usage customers, creating pressure
to introduce consumption controls.

The response is observable through a sequence of publicly documented
pricing changes. OpenAI and Anthropic moved API customers to prepaid
credit balances with drawdown, auto-recharge, and expiration [2, 8].
GitHub Copilot, a seat product, added consumptive "premium request"
billing in mid-2025 [6]. Cursor repriced in June 2025 and publicly
clarified within two weeks [3, 4]. AI agent products went further
and priced the outcome itself: Intercom's Fin charges per resolved
conversation, Salesforce's Agentforce per conversation [7, 9]. Data
platforms formalized consumption credits, now extended to AI workloads
[10].

Pricing practice has changed substantially. A systematic, reproducible
account of the pricing primitives deployed by AI companies and
AI-selling cloud platforms remains limited: which primitives they
converged on, how the mechanics differ, and what the convergence demands
from the engineering of billing systems. The relationship runs in both
directions: the pricing designs a company chooses determine what its
metering, entitlement, and settlement systems must do, and the
capabilities of those systems bound which pricing designs are practical
to operate. This paper supplies that account with a sampling method
designed for the question rather than inherited from any single ranking.
The research questions are:
\begin{itemize}
\item \textbf{RQ1:} What pricing primitives did AI-native companies and cloud platforms selling AI deploy between 2023 and 2026, and how do they distribute across the AI-native and cloud-platform cohorts?

\item \textbf{RQ2:} What commitment and consumption rules accompany these pricing primitives?

\item \textbf{RQ3:} What implications do the observed pricing models have for the design of billing platforms?
\end{itemize}

Scope of the claims. All prevalence figures in this paper are
complete-case descriptions of publicly observable pricing pages at the
September 16--17, 2026 capture: they describe what observable pages
show, not what all AI companies do, and 23 of the 52 AI-native rows
publish no pricing (Section 3, O6; Section 8). The historical panel
(Fig. 1) is a set of repeated archived cross-sections whose observable
composition changes between checkpoints, not a longitudinal trend. The
architecture briefs of Section 5 are evidence-informed design
propositions derived by engineering reasoning from the observed
constructs, not validated architectures or evaluated implementations.
The projections of Section 7 are falsifiable expectations, not findings.

\begin{figure}[h]
\centering
\includegraphics[width=\linewidth]{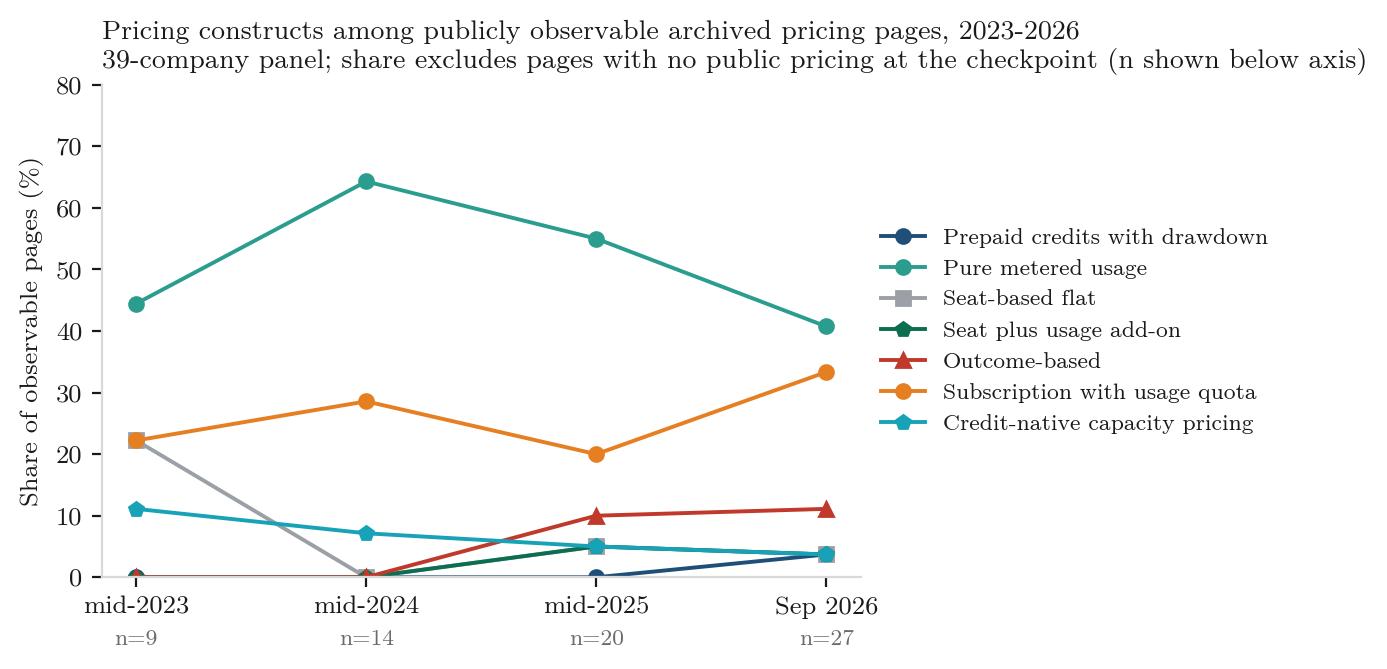}
\caption{Repeated cross-sections of a fixed 39-company historical panel: pricing-construct shares among publicly observable archived pricing pages at four checkpoints, mid-2023 to September 2026. The set of observable companies differs across checkpoints (n = 9, 14, 20, 27 observable pages), so shares across checkpoints are not directly comparable and no trend is estimated; straight segments connect measured points only, with no smoothing or extrapolation. Points are measured shares of pages with public pricing at each checkpoint. Seven companies are observable at all four checkpoints, and within that balanced subset exactly one construct transition occurs (Gamma, seat-flat to subscription with usage quota, between mid-2023 and mid-2024). Panel membership is drawn from companies notable in 2026 (survivorship bias); each point is a page's state at an archive capture date, not a dated change event; pages are marketing surfaces, not contracts. Source: replication dataset, Historical panel tab (archived checkpoints) and Companies tab (September 2026 point).}
\end{figure}

This study makes three contributions. First, it provides a source-traceable
dataset of publicly observable AI pricing designs across AI-native
companies, cloud platforms, and a purposive set of distinctive pricing
models. Second, it develops a taxonomy that distinguishes pricing
primitives from the mechanics governing credits, quotas, consumption,
and outcomes, including a sensitivity analysis of hybrid products.
Third, it translates the observed pricing structures into requirements
for automated, configurable billing platforms, showing how different
economic primitives imply different state, metering, entitlement, and
attribution mechanisms.

\hypertarget{study-design}{%
\section{Study design}\label{study-design}}

\hypertarget{sampling-method}{%
\subsection{Sampling method}\label{sampling-method}}

The sample is designed from the research question, not anchored to any
single ranking. Cohort 1 (AI-native core): all 50 companies on the
Forbes AI 50 (2026) list [25]; a company whose pricing is not
publicly observable stays in the sample coded as such rather than being
substituted. Cohort 2 (cloud platforms selling AI): any cloud or data
platform with a managed AI service and a public AI-specific pricing
page; membership is determined by this rule rather than ad hoc
selection, and the enumeration is recorded row-level in the dataset.
Because no authoritative exhaustive population exists for this category,
the study does not claim that the resulting cohort represents all
qualifying platforms. Cohort 3 (purposive): 10 candidate companies with
distinctive pricing designs, frozen before analysis; this cohort feeds
design analysis only and never enters prevalence percentages. Row
reconciliation: the 50 Forbes companies yield 52 AI-native product rows
(OpenAI and Anthropic are each coded on two product lines, API and
consumer app); the purposive cohort contributes 9 rows (one frozen
candidate, ElevenLabs, appeared on the Forbes AI 50 and moved to the
core cohort under the code-as-is rule); with 14 cloud-platform rows, the
dataset has 75 rows (Fig. 2). Cohort membership is mutually exclusive after this
reconciliation. My employer is excluded (see Section 2.5). Study window:
2023 to the September 16--17, 2026 capture date, covering the commercial
generative-AI repricing wave. Detailed inclusion, exclusion, and
stopping rules are in Appendix A.

\begin{figure}[H]
\centering
\includegraphics[width=\linewidth]{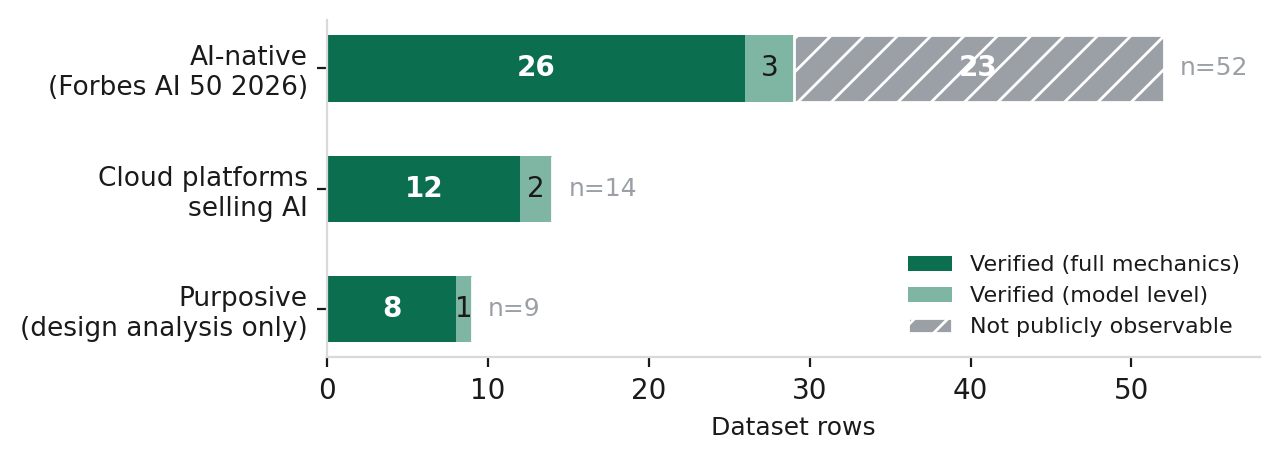}
\caption{Cohort composition and public observability across the 75 dataset rows (captured September 16, 2026). Source: replication dataset, Companies tab.}
\end{figure}

\hypertarget{data-sources}{%
\subsection{Data sources}\label{data-sources}}

Primary sources only: live pricing pages, official documentation, and
official changelogs or pricing-announcement posts - no third-party
summaries or analyst reports. For the historical panel, Wayback Machine
snapshots of the same companies' pricing pages are coded at three
archived checkpoints (mid-2023, mid-2024, mid-2025); the fourth
checkpoint (September 2026) is the live September 16--17 capture; because
archive coverage varies by page, the checkpoints form repeated
cross-sections of a fixed 39-company panel rather than a balanced
longitudinal sample (Section 7); archive coverage and gaps are logged
per page in the dataset (Wayback coverage tab).

\hypertarget{coding-schema}{%
\subsection{Coding schema}\label{coding-schema}}

Each row codes one company on its pricing primitive, credit/usage
mechanics (minimum commitments, drawdown order, recharge, expiry,
rollover), unit of charge, entry price, and - for subscription with
usage quota rows - second-level quota structure. Operational
definitions, decision rules, coding-unit conventions, and boundary cases
are in Appendix A and recorded in the dataset (Schema and Codebook
tabs). Two rules matter most for hybrid products. First, the coding unit is the anchor SKU, the plan the typical self-serve buyer actually purchases, with secondary models recorded in notes. Second, seat-plus-usage (P4) is coded when buyers purchase seats first and usage is the add-on, such as premium requests or extra usage dollars; when usage instead draws on an included quota that resets inside a flat tier, the row is coded as subscription with usage quota (P6). Rows where both readings are defensible carry an alternate code (Section 3 and Appendix A.2). I designed the taxonomy and performed all primary coding.

\hypertarget{reproducibility}{%
\subsection{Reproducibility}\label{reproducibility}}

The dataset is the replication package: every row carries its source URL
and capture date. An independent coder re-coded a blinded random 20\%
subset (15 of 75 rows, stratified by cohort; seeded and reproducible,
protocol in the replication dataset). Agreement with the original
coding: observability status 15/15. Of the 15 audited rows, 4 were not
publicly observable and therefore had no pricing-primitive code; among
the remaining 11 observable rows, pricing-primitive agreement was 8/11
(72.7\%). Cohen's $\kappa$ was 0.52 when calculated on the 11 observable
rows and 0.71 when observability status was included as an additional
coding category. Given the small audited subset, these agreement
statistics are an audit of coding consistency rather than a precise
estimate of population-level inter-coder reliability. Because I both
designed the taxonomy and applied the primary coding, the audit measures
whether an independent coder applying the same written codebook
reproduces my codes; it validates neither the taxonomy's partition
itself nor the 80\% of rows outside the audited subset. All three
primitive disagreements (Notion, Vercel, Salesforce Agentforce) were
adjudicated as boundary cases and now carry primary and alternate codes
(Codebook tab, section 4); no original code was overturned on the
captured evidence. After written Stripe approval, the replication
dataset will be published with the paper to support external audit. Until it is released, reproducibility is limited to re-coding from the cited public pages using the codebook rules described in this paper; the figures cannot yet be regenerated by outside readers. The
planned package includes the full codebook, primary and alternate codes
for boundary rows, Wayback coverage logs, the historical panel, and the
analysis and figure-generation code, so external re-coding and figure
regeneration are straightforward. The same package supports
evidence-traceability for the derived platform requirements: each
architecture brief in Section 5 and Table 2 cites the observed
constructs and mechanics it draws on; the requirements themselves are my
engineering interpretation, not coded observations. The snapshot-freeze
and update policy is in Appendix A. The pricing codes were assigned manually by me, not by an AI system. AI coding agents assisted in
building the research workflow and analysis code under my direction; I
defined the methodology, reviewed and validated all evidence, and take
full responsibility for the analysis and conclusions.

\hypertarget{ethics-positionality-and-conflict-boundary}{%
\subsection{Ethics, positionality, and conflict
boundary}\label{ethics-positionality-and-conflict-boundary}}

All artifacts are public marketing or documentation surfaces; every
coded field carries a source URL and capture date, and nothing an
outsider could not see enters the dataset. I work at a billing platform.
To limit potential conflicts and support reproducibility, the study
excludes my employer from the sample and uses only publicly available
artifacts. No private customer, product, or operational information was
used. The same practitioner background shapes the taxonomy itself: the
construct definitions, and the separation of primitives from mechanics,
reflect a billing-platform builder's decomposition of the pricing space,
and a researcher from another tradition might carve it differently; the
written codebook, the dual-code convention, and the independent-coder
audit are the structural mitigations.

\hypertarget{what-companies-deploy-a-taxonomy-of-pricing-primitives}{%
\section{What companies deploy: a taxonomy of pricing
primitives}\label{what-companies-deploy-a-taxonomy-of-pricing-primitives}}

Dataset: 75 rows coded (captured September 16, 2026): 52 AI-native product rows,
14 cloud platforms selling AI, 9 purposive. 46 offerings verified with
full mechanics; 23 confirmed not publicly observable. For six offerings,
public sources confirmed the pricing model but not all plan-level
mechanics: rate values or complete rate schedules for the AI charge unit
(Sierra per-outcome, Zendesk AI per-resolution, Datadog LLM
Observability, Nebius), per-tier usage-limit detail (ChatGPT consumer),
and web-published prices (Speak, app-store distribution); gaps are
recorded row-level in the dataset. The coded entry-price field,
populated for 44 of the 52 observable rows across all three cohorts, shows a free entry point
(free tier, trial, or starter credits) on 25 rows, pay-as-you-go or
on-demand entry on 13 rows, a paid entry of \$8--\$20 per month on five
rows, and per-resolution entry at \$0.99 on one row. Percentages below
are computed over observable rows only.

P1 - Prepaid credits with drawdown. The customer purchases a balance up
front and consumes it; mechanics include auto-recharge, expiration, and
balance limits. Observed: OpenAI API [8], Anthropic Claude API
[2].

P2 - Pure metered usage. Pay per unit consumed, billed in arrears. The
classic cloud model; expected across model-API and infrastructure
providers.

P3 - Seat-based flat. Per user per month, unmetered. The legacy SaaS
default, under pressure everywhere marginal cost is real.

P4 - Seat plus usage add-on. A seat carries an allowance; overage bills
consumptively. Observed: GitHub Copilot premium requests [6], Cursor
plan usage allowances [3, 4].

P5 - Outcome-based. The unit of charge is a business outcome, not a
resource. Observed: Intercom Fin at \$0.99 per resolution with a
50-resolution monthly minimum [7], Salesforce Agentforce per
conversation [9].

P6 - Subscription with usage quota. A flat tier with metered caps, often
denominated in credits. Observed: Claude consumer plans [1],
ElevenLabs monthly credit quotas [5].

P7 - Credit-native capacity pricing. Capacity is sold in
provider-denominated credits with published consumption-rate schedules,
sitting between P1 and P2: credits as the unit of capacity rather than a
prepaid wallet. Observed once in the dataset: Snowflake's consumption
credits, extended with AI-specific credit consumption [10]; a single
observation supports no prevalence claim. Five terminology notes apply.
A pricing primitive is the primary packaging and economic unit used to
charge the customer; pricing mechanics are the operational rules
governing commitment, consumption, rollover, expiration, recharge, and
settlement. Settlement timing is treated as a secondary mechanic, not a
distinct primitive (see O2). Credit denominations serve three
economically distinct functions in the dataset: prepaid balances
purchased before consumption (P1), provider-defined units included in a
subscription allowance (P6), and provider-defined capacity units (P7);
treating them as equivalent would obscure the difference between
payment, entitlement, and resource measurement. A pricing layer is one
composable dimension of a hybrid product (packaging anchor, allowance or
payment mechanism, billable unit). Quota terms are used consistently
throughout: an allotment is a fixed included quantity per period; a
refreshing allowance renews at the start of each reset window; rollover
carries unused allowance into the next window.

\begin{figure}[H]
\centering
\includegraphics[width=\linewidth]{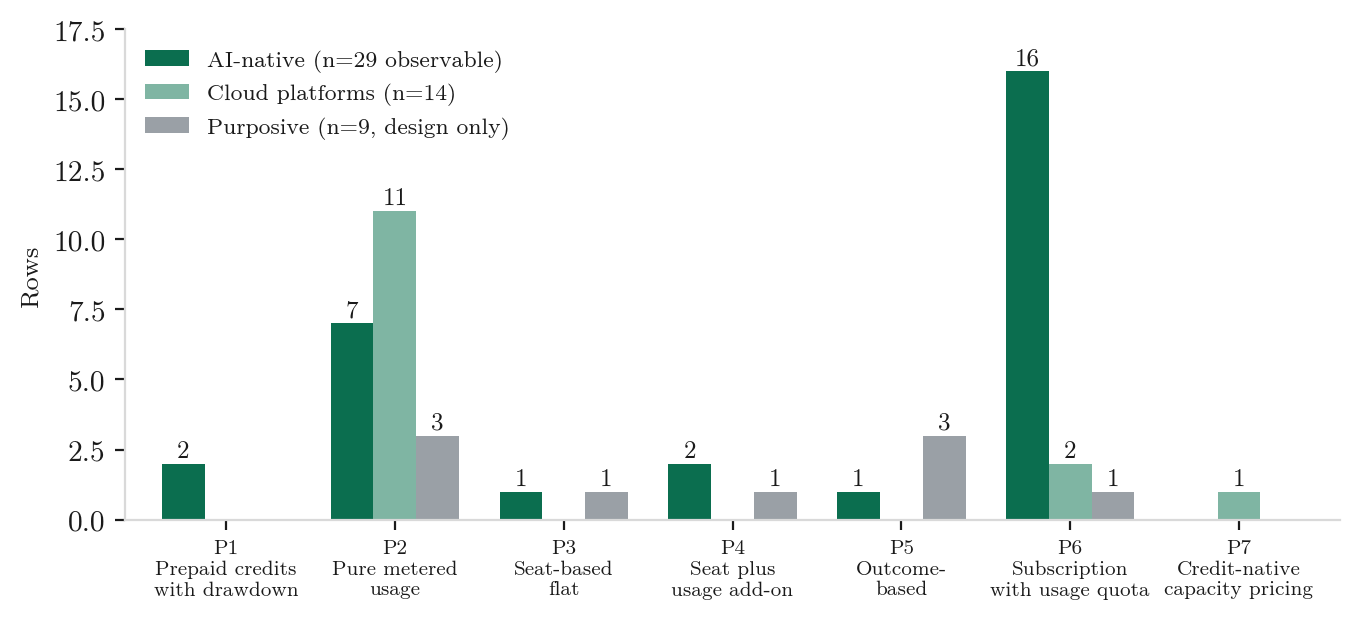}
\caption{Pricing-primitive distribution by cohort, observable rows only (AI-native n=29; cloud platforms n=14; purposive n=9). Source: replication dataset, Companies tab.}
\end{figure}

\hypertarget{findings}{%
\subsection{Findings}\label{findings}}

All prevalence percentages below are computed over observable rows only
and use the primary pricing code under the anchor-SKU coding rule;
alternate codes for hybrid boundary products are reported with the
sensitivity analysis, and the layered reading is a co-equal result, not
a caveat (Fig. 3).

O1: Subscription with usage quota (P6) is the dominant observable model
among AI-native companies: 16 of 29 observable AI-native rows (55\%) - a
complete-case share over pricing-observable rows; 23 of the 52 AI-native
rows publish no pricing (O6), so the corresponding share over all
Forbes-listed companies is unknown (Section 8). Among consumer-facing AI
apps it is the modal pattern - Runway, Synthesia, HeyGen, Suno, Gamma,
Krea, ElevenLabs, Lovable, Perplexity, Notion, and Midjourney all sell
flat tiers metered in credits, compute units, or GPU-time. Synthesia's
pricing page states the mechanism outright: "Credits are now the shared
currency across all AI usage-based features" - company marketing copy,
cited as evidence of how the offer is publicly represented, not of the
implemented system.

O2: Model APIs split on settlement timing. The comparison set is the
verified AI-native rows whose anchor SKU is a model API or AI inference
infrastructure: OpenAI, Anthropic, Cohere, Together, Baseten, Fireworks,
Fal, and Black Forest Labs (8 rows). Databricks (metered in
provider-specific DBUs rather than model-API units) and Mistral (whose
anchor SKU is a subscription bundling API credits, with pay-as-you-go
API billing as a secondary rail) sit on the boundary and are excluded
under the anchor-SKU rule. Within this set, OpenAI and Anthropic require
prepaid credit drawdown; the other six bill postpaid. The unit of charge
is not uniform: OpenAI, Anthropic, and Cohere bill per token, while the
other five also meter GPU-time or per-second output units.

O3: In this sample, outcome pricing is real but narrow, and it lives
exactly where outcomes are contractible - observable and auditable
enough to serve as a contractual billing event. Only 1 of 52 AI-native
rows states outcome pricing (Sierra - publicly, without a rate card). In
the purposive cohort all three outcome-priced companies (Agentforce,
Fin, Zendesk AI) sell into customer service, where a "resolution" is
observable, attributable, and contractible. Outside customer service, no
outcome-priced offering was observed in the sample.

O4: In this sample, established seat products absorb AI cost by layering
meters on top of seats, not by repricing the seat. GitHub Copilot adds
consumptive premium-request allowances; Microsoft bundles Copilot into
M365 seats (plans with Copilot built in were listed at \$18--32 per
user/month at the September 2026 capture); Notion moved AI to credit-based
pricing on top of seat plans.

O5: Similar packaging anchors can conceal materially different operating
mechanics. Expiry (Adobe free credits expire monthly; OpenAI credit
expiration applies), rollover (HeyGen allows it, Midjourney fast-time
does not), minimums (Fin's 50 resolutions/month; Pinecone's \$50/month),
recharge policy (OpenAI auto-recharge), and even time-of-day pricing
(DeepSeek off-peak discounts, unique in the sample) carry the
differentiation. Unlike O1-O4, which rest on row counts, O5 is an
illustrative observation of mechanics variety; the time-of-day case is a
single observation, not a prevalence finding.

O6: Pricing transparency tracks go-to-market motion. 23 of the 52
AI-native rows (44\%) publish no pricing at all; every cloud-platform
row is observable (14/14). Coding the purchase path on each company's
public site (dataset column M; captured September 16, 2026): of the 23
non-observable rows, 12 (52\%) show an explicit demo or contact-sales
call to action, 2 are partnership-driven, 4 offer self-serve signup
without published rates, and 5 are unclear (per-company purchase-path
codes: dataset column M). Of the 52 observable rows across the full
dataset, 45 are pure self-serve, 6 pair published rates with a sales
call to action, and the remaining row (Sierra) publishes its pricing
model but shows no purchase call to action. Within the coded public
purchase paths, non-observable pricing was more frequently accompanied
by a demo or contact-sales call to action than observable pricing. This
is an association, not proof of the underlying sales motion: the coding
reflects homepage calls to action, not the company's actual sales
motion. Limitations: single-coder coding of public CTAs on September 16,
2026; Surge AI's site was unreachable at capture; the "unclear" rows are
mostly research-stage labs with no public purchase path.

O7: Cohorts differ in dominant pricing structure, consistent with differing cost structures. Cloud platforms selling AI
generally expose AI workloads through consumption-oriented pricing
structures (12 of 14: per-token, per-GPU-hour, credits, or abstract
units like Cloudflare Neurons); AI-native consumer apps sell capacity
quotas; agents price outcomes; devtools keep seats and add usage
allowances (Cursor, Cognition's Devin and Windsurf offerings).
Cognition's row is a boundary case, coded P6 primary with a P4
seat-plus-usage alternate (see the sensitivity analysis below and the
replication dataset, Codebook tab). The link to cost structure is an
interpretation: the study codes public prices, not providers' costs, so
the association is descriptive and untested.

Sensitivity. Nine rows sit on coding boundaries (Perplexity, Replit,
Devin/Windsurf, Datadog LLM Observability, Zendesk AI, Sierra, from the
September 17 re-verification; Notion, Vercel, and Salesforce Agentforce,
added after the second-coder audit). Each is coded with a primary and an
alternate defensible code (replication dataset, Codebook tab, section
4). Recomputing the primitive distribution under the alternates - the
minimum and maximum over all 2\^{}9 = 512 assignments of the nine
boundary rows' primary and alternate codes, with Sierra's alternate
removing it from the observable denominator - moves P6's AI-native share
only within 45--61\% (13--17 of 28--29 rows) and leaves it the largest
primitive in every variant; O2, O3, O6, and O7 are unchanged. Varying the
product-line coding and the boundary rows together, the P6 share ranges from 11 of 27
(41\%) to 17 of 26 (65\%) across all joint variants, and P6 remains the
largest primitive in each: its lowest count (11) still exceeds the next
largest primitive count (pure metered usage, 7 rows).
One limit applies. The sensitivity analysis bounds the rows known to be
ambiguous; it does not bound unknown coding error. The coding
unit is the product line, so OpenAI and Anthropic each contribute two
rows (API and consumer app); collapsing each company to one row under
either product-line choice yields 27 companies with subscription with
usage quota at 14--16 (52--59\%), still the largest share, so no finding
rests on the two-row coding. Both codings are recorded in the dataset so
external re-coding can test them.

The audit also exposed a limitation of the taxonomy. The three primitive
disagreements did not dispute prices or mechanics; they assigned primacy
to different layers of the same hybrid product: the packaging anchor (a
per-seat subscription at Notion and Vercel), the allowance or payment
mechanism (prepaid Flex Credits at Salesforce Agentforce), and the
economic billable unit (a completed conversation). I read this as
evidence that hybrid products combine separable pricing layers, and that
a single-label taxonomy supports prevalence analysis only by forcing one
layer into the primary cell. The dual-code convention records the
layering instead of hiding it.

\hypertarget{timeline-of-public-repricing-events-20232026}{%
\section{Timeline of public repricing events,
2023--2026}\label{timeline-of-public-repricing-events-20232026}}

This timeline (Fig. 4) is built from dated official announcements - blog posts,
press releases, changelog entries, and release notes - captured as of
September 16--17, 2026; each event's source URL and retrieval context
are recorded in the replication dataset (Timeline events tab), the
source ledger for this section. It is announcement-based evidence: it
records when companies publicly changed AI pricing, not a complete
market history. For nine anchor companies, announcement dates are
distinguished from the earliest archived observation of the relevant
pricing page (reconstruction method and per-company results in Appendix
A and the Wayback coverage tab of the replication dataset); where the
earliest archived snapshot postdates the announcement, archive evidence
corroborates but cannot date the change.

Feb 22, 2023 - Notion opens Notion AI to everyone as a paid add-on on
its established SaaS product.

Mar 14, 2023 - Duolingo launches Duolingo Max, a premium GPT-4 tier
above its Super subscription.

Mar 14, 2023 - Intercom launches Fin, its AI support agent, with outcome
pricing at \$0.99 per resolution.

Jul 18, 2023 - Microsoft announces Microsoft 365 Copilot pricing: \$30
per user per month as an AI add-on to its established SaaS suite.

Sep 13, 2023 - Adobe ships the AI-powered Creative Cloud release, adding
generative-credit allowances to established subscriptions.

2023--2024 - OpenAI documents prepaid API billing: the help article is
archived from August 23, 2023 with a prepaid-billing slug but is undated
[8], and a developer-community thread begun February 28, 2024
reports the switch for API usage [26]; the archive establishes
documentation by August 2023, not the policy's effective date.

Aug 28, 2024 - Zendesk announces outcome-based pricing for AI agents,
charging per automated resolution.

Sep 3, 2024 - Canva reprices its Teams subscriptions citing AI features
- an established SaaS product repricing around AI.

Sep 18, 2024 - HubSpot launches Breeze Intelligence with credit-based
data-enrichment pricing on its established CRM.

May 15, 2025 - Salesforce introduces flexible Agentforce pricing (Flex
Credits, per-action consumption) alongside the original per-conversation
model.

June 16, 2025 - Cursor ships a usage-based repricing of Pro and
introduces Ultra; following confusion, it publishes a clarification on
June 30.

June 18, 2025 - GitHub updates Copilot's consumptive billing experience
for premium requests [6].

July 28, 2025 - Anthropic announces weekly rate limits on Claude Pro and
Max subscriptions, aimed at Claude Code power users (effective Aug 28,
2025; applies to \textless5\% of subscribers; Max users may buy extra
usage at standard API rates) [24].

Aug 18, 2025 - Synthesia replaces minutes with credits on self-serve
plans, changing the unit of consumption.

Feb 24, 2026 - Notion launches Custom Agents in free public beta and
announces credit-based pricing, effective May 4, 2026.

Apr 6, 2026 - Snowflake's \texttt{AI\_SERVICES} billing breakout reaches GA,
itemizing consumption across implemented AI credits services.

One expected entry has no verifiable public date and is recorded as
incomplete evidence rather than dated: Anthropic's prepaid-credits
requirement for API usage (official support article, undated). It is
excluded from dated counts; the earliest archived snapshot of the
support article is September 18, 2025, which anchors the page state but
not the change date.

Within this window, a distinct cohort is established SaaS companies
moving into AI monetization: Notion (paid add-on in 2023, credits in
2026), Duolingo, Intercom, Microsoft, Adobe, Zendesk, Canva, HubSpot,
Salesforce, and GitHub. The dominant pattern is layering, not migration:
AI arrives as an add-on or tier on top of seat subscriptions, and
consumption mechanics - credits, usage, or outcomes - attach where AI
compute cost is real. Only one company in this cohort, Notion, has
publicly moved an AI offering from a flat add-on to credits. Companies
whose AI announcements carry no verified pricing change stay out of this
cohort; product messaging alone is not treated as repricing.

Across the dated events, repricing often happens at product velocity,
in public, with community feedback following the change; this is a
reading of a small set of announcements, not a measured rate. Cursor's
two-week ship-then-clarify cycle is the clearest example [3, 4]. The timeline contains five dated events in 2023, four in 2024 (one
community-dated), five in 2025, and two through April 2026; because 2026
is right-censored, these counts are not directly comparable across
complete calendar years and are not evidence of an increasing annual
cadence.

\begin{figure}[H]
\centering
\includegraphics[width=\linewidth]{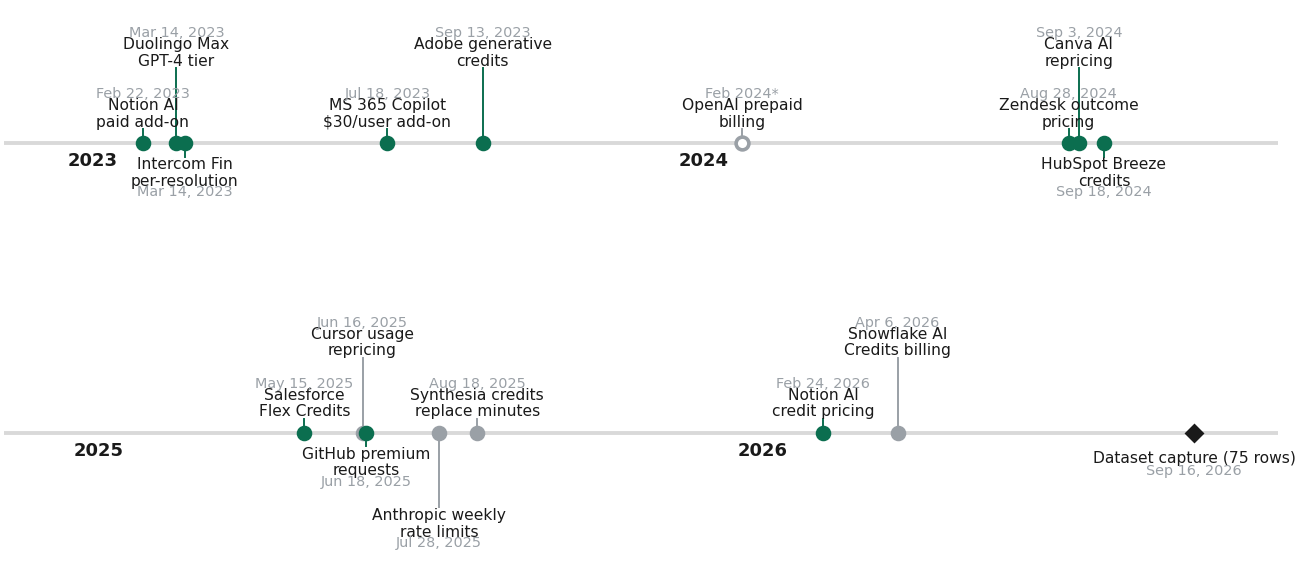}
\caption{Dated public repricing events from official announcements, 2023--2026 (captured September 16--17, 2026). Green markers: established SaaS companies adding AI monetization; gray markers: AI-native or cloud-platform repricing (the OpenAI marker is community-dated). Source: replication dataset, Timeline events tab.}
\end{figure}

\hypertarget{implications-for-billing-platform-design}{%
\section{Implications for billing-platform
design}\label{implications-for-billing-platform-design}}

This section derives evidence-informed design propositions from the
findings through engineering reasoning: grounded in the observed
constructs, they are not validated architectures or evaluated
implementations. Each construct gets an architecture brief in the same
pattern: evidence, design action, and the main risk plus the customer
control. Capability details are consolidated in Table 2 rather than
repeated in prose. For traceability, the observed AI-native rows behind
the Table 2 constructs number, of 29 observable rows: prepaid credits 2,
pure metered usage 7, seat-based flat 1, seat plus usage add-on 2,
outcome-based 1, and subscription with usage quota 16; credit-native
capacity is observed in one cloud-platform row (Snowflake). These counts
are small for several constructs, so the briefs are design propositions
rather than prevalence-weighted recommendations.

\hypertarget{prepaid-credits-with-drawdown}{%
\subsection{Prepaid credits with
drawdown}\label{prepaid-credits-with-drawdown}}

Evidence. Rare in the sample (2 of 29 observable AI-native rows:
OpenAI's API and Anthropic's Claude API). Captured OpenAI
prepaid-billing mechanics: balance purchase up front, auto-recharge,
credit expiration, balance limits.

Action. Own a balance ledger of credit lots (source, amount, expiration,
drawdown order) with atomic, idempotent debits and threshold-triggered
automatic recharge. Expose the spend controls: configurable balance
limits, manual and automatic top-up, low-balance alerts, and defined
exhaustion behavior. The balance is the customer's budget boundary,
fixed before consumption rather than discovered on an arrears invoice.

Risk. This introduces a ledger and balance-management problem that is
distinct from ordinary usage metering: double-spend races, unnoticed
expiry, and ambiguous refunds are ledger failure modes, not metering
bugs. Expiration raises accounting-treatment and consumer-protection
considerations where unused balances are forfeited; depending on the
contractual and accounting treatment, forfeited credits may also have
breakage implications.

\hypertarget{pure-metered-usage}{%
\subsection{Pure metered usage}\label{pure-metered-usage}}

Evidence. The largest single share in the historical panel (41\% of
observable pages, September 2026) and the cloud-platform default (11 of
14 rows): per-token with input/output splits, per-GPU-hour, per-GB,
per-event, time-of-day tiers (DeepSeek), batch discounts (AWS Bedrock),
provisioned-throughput overlays (AWS, Azure).

Action. Own effective-dated, versioned rate cards over an idempotent
event stream (dedup keys, late-event windows), with dimension versioning
so rate splits evolve as data, not code. Expose budget alerts, caps, and
anomaly detection - the model carries no inherent spend boundary.

Risk. Dropped events are silent revenue leakage; double-counted events
become disputes. Metering accuracy is the invoice.

\hypertarget{seat-based-flat-pricing}{%
\subsection{Seat-based flat pricing}\label{seat-based-flat-pricing}}

Evidence. Rare among AI-native sellers (Speak; prices not public) but
the established-SaaS default: at the September 2026 capture, Microsoft 365
Copilot is bundled into seat plans listed at \$18--32 per user/month (the
July 2023 launch price of \$30 per user per month as an add-on is a
dated event in Section 4, not a current price).

Action. Own the entitlement map (user to plan to features), provisioning
lifecycle, and proration; no metering path exists.

Risk. Unmetered AI cost per seat can subsidize power users into
unprofitability - one source of pressure motivating the other constructs
observed in this study. Seat-count reconciliation drifts against the
customer's identity system.

\hypertarget{seat-plus-usage-add-on}{%
\subsection{Seat plus usage add-on}\label{seat-plus-usage-add-on}}

Evidence. GitHub Copilot (monthly premium-request allowance, consumptive
overage), Cursor (plan usage limits, Ultra at 20x Pro), Replit (included
usage dollars above subscription face value).

Action. Compose two rating paths - seat entitlement plus consumptive
allowance with arrears overage - and settle both on one invoice. Make
the allowance-versus-overage boundary explicit to the customer in real
time, with overage caps.

Risk. Boundary ambiguity fails publicly: Cursor's June 2025
ship-then-clarify episode [3, 4]. Dispute-handling cost grows with
the number of rating paths, not customers.

\hypertarget{outcome-based-pricing}{%
\subsection{Outcome-based pricing}\label{outcome-based-pricing}}

Evidence. Customer service only in this sample (Sierra; Intercom Fin at
\$0.99 per resolution with a 50-resolution monthly minimum; Salesforce
Agentforce; Zendesk outcome positioning).

Action. Own an outcome registry with versioned definitions and an
attribution pipeline where every billable event carries its definition
version and customer-visible evidence. Support contingent recognition as
an accounting policy, and expect minimums (Fin's 50 resolutions) to
hybridize the model with subscription.

Risk. Definition drift, mis-attribution disputes, and gaming pressure on
the detection rule. Unless the outcome is defined, detected, and
auditable, the invoice is not defensible.

\hypertarget{subscription-with-usage-quota}{%
\subsection{Subscription with usage
quota}\label{subscription-with-usage-quota}}

Evidence. The largest AI-native share (16 of 29 observable rows, 55\%)
and present at every checkpoint of the historical panel's repeated
cross-sections (Fig. 1); within the seven-company balanced subset,
construct codes are stable across three years except a single transition
- Gamma, into this construct, between mid-2023 and mid-2024 (Section 7).
Internally these 19 products (16 AI-native, 2 cloud-platform, 1
purposive) differ: captured pages specify the quota structure for 9
rows, and the other 10 are not inferred (Table 1): shared credit pools
across features (five: Clay, Runway, Synthesia, Krea, Lovable), quota
with metered overage (three: Vercel, Supabase, Cognition), time-windowed
allowances (Cognition), a metered quota with an unmetered rail
(Midjourney), and one separate per-feature limit (Synthesia's video
minutes alongside its shared pool). No captured page presents a complete
per-feature quota schedule - a different included limit for every
feature - under one fixed fee.

Table 1. Observed quota structures within subscription with usage quota
(19 rows). Coded only from captured pricing-page text (September 16,
2026); rows can combine structures, so category counts sum to 21
assignments across 19 unique products: Cognition appears under both
quota + metered overage and time-window quotas, and Synthesia under both
shared pool and separate per-feature quotas. Source: replication
dataset, Companies tab columns N-O.

{\footnotesize\setlength{\tabcolsep}{3pt}%
\begin{longtable}[]{@{}p{0.24\linewidth}p{0.06\linewidth}p{0.42\linewidth}p{0.22\linewidth}@{}}
\toprule
\textbf{Quota structure} & \textbf{Rows} & \textbf{Companies} &
\textbf{Captured example}\tabularnewline
\midrule
\endhead
Shared pool across features & 5 & Clay, Runway, Synthesia, Krea, Lovable
& Lovable: credits measure build, hosting, and AI usage from one
balance\tabularnewline
Quota + metered overage & 3 & Vercel, Supabase, Cognition & Vercel: \$20
included usage credit, then metered overage\tabularnewline
Time-window quotas & 1 & Cognition & Devin/Windsurf: daily plus weekly
allowances\tabularnewline
Quota + unmetered rail & 1 & Midjourney & fast-GPU-time quota plus
unmetered Relax Mode\tabularnewline
Separate per-feature quotas & 1 & Synthesia (with shared pool) & shared
credits plus a distinct video-minute limit\tabularnewline
Unknown/unspecified & 10 & ChatGPT, Claude, Mistral, Perplexity, Notion,
Gamma, ElevenLabs, HeyGen, Suno, Adobe Firefly & captured page does not
state the structure\tabularnewline
\bottomrule
\end{longtable}}

Action. Build the structures of Table 1 as first-class, configurable
capabilities - quota definition, entitlement enforcement, shared and
per-feature balances, periodic and windowed resets, overage transitions,
mixed metered and unmetered rails - and keep the entitlement gate and
the settlement meter on separate paths. Expose quota-consumption
visibility and near-exhaustion alerts.

Risk. Gate/ledger conflation produces revenue leakage or user-facing
surprises; resets race at window boundaries; rollover and expiry
differences (HeyGen rolls over, Midjourney does not) are where customers
get surprised. Ten of 19 pages do not disclose the structure, so the
platform must tolerate unspecified composition. This is design
interpretation; the counts are the measured evidence.

\hypertarget{credit-native-capacity}{%
\subsection{Credit-native capacity}\label{credit-native-capacity}}

Evidence. One company (Snowflake): capacity sold denominated in provider
credits with defined consumption rates, between prepaid wallet and
metered usage.

Action. Own published consumption-rate schedules per resource and credit
metering against balances.

Risk. The provider can rebalance consumption rates behind the credit,
shifting rate-opacity risk to the buyer. A single observation supports
no prevalence claim.

\hypertarget{cross-cutting-pricing-change-velocity}{%
\subsection{Cross-cutting: pricing-change
velocity}\label{cross-cutting-pricing-change-velocity}}

Evidence. Public repricing moves at product velocity: a repricing
shipped June 16 was clarified June 30 [3, 4].

Action. Treat pricing as data: rate-card versioning (\S 5.2) turns
a repricing from a migration into an edit. The 2025 repricing cadence
is, from the platform side, an argument for configuration-driven rating
across every construct above.

Table 2. Synthesis: the platform state, first-class controls, primary
failure mode, and customer spend control each pricing construct demands
(Sections 5.1-5.7).

{\footnotesize\setlength{\tabcolsep}{3pt}%
\begin{longtable}[]{@{}p{0.04\linewidth}p{0.16\linewidth}p{0.21\linewidth}p{0.21\linewidth}p{0.13\linewidth}p{0.15\linewidth}@{}}
\toprule
\textbf{\#} & \textbf{Construct} & \textbf{Core platform state} &
\textbf{First-class controls} & \textbf{Primary failure mode} &
\textbf{Customer spend control}\tabularnewline
\midrule
\endhead
1 & Prepaid credits with drawdown & balance ledger of credit lots &
atomic drawdown, expiry processing, auto-recharge & double-spend races;
unnoticed expiry & balance limits, low-balance alerts\tabularnewline
2 & Pure metered usage & versioned rate cards + event stream &
idempotent ingestion, dimension versioning & dropped or double-counted
events & none inherent; platform budgets and caps\tabularnewline
3 & Seat-based flat & entitlement map & provisioning lifecycle,
proration & seat-count drift; unmetered AI cost & per-seat
predictability\tabularnewline
4 & Seat plus usage add-on & two composed rating paths & explicit
allowance/overage boundary & boundary ambiguity (Cursor 2025) & live
allowance-vs-overage split\tabularnewline
5 & Outcome-based & outcome registry + attribution pipeline & definition
versioning, event evidence & definition drift, mis-attribution &
counted-event transparency\tabularnewline
6 & Subscription with usage quota & quota registry: scope, window,
overage, rollover & separated gate and meter, resets & gate/ledger
conflation; reset races & quota visibility, near-exhaustion
alerts\tabularnewline
7 & Credit-native capacity pricing & consumption-rate schedules & credit
metering against balances & rate rebalancing behind the credit &
consumption-rate transparency\tabularnewline
\bottomrule
\end{longtable}}

\hypertarget{automation-opportunities}{%
\subsection{Automation opportunities}\label{automation-opportunities}}

The architecture briefs above are requirements observed from public
pricing designs. The automation opportunities below are proposed
engineering responses to those requirements; none is an observed result,
and none is evaluated here.

Configuration-as-code for pricing artifacts. The repricing velocity
documented in Section 4, including the ship-then-clarify pattern, argues
for representing rate cards, quota schedules, and credit consumption
tables as versioned, declarative configuration rather than code paths,
so that a repricing is a reviewed data change with automated schema
validation rather than a deploy.

Automated consistency checks and state invariants. Each construct
implies checkable invariants: prepaid ledgers conserve value across lots
(granted equals consumed plus expired plus remaining balance); quota
subscriptions require agreement between the entitlement gate and the
settlement meter at every reset boundary; outcome pricing requires every
billable event to carry a registry definition version. These invariants
are candidates for continuous automated verification rather than
periodic manual reconciliation.

Versioned rollout and rollback of pricing changes. Rate-card versions
with explicit effective windows allow automated rollout, rollback, and
replay of recorded usage against a candidate rate card before
publication - a direct response to the observed change velocity.

Automated attribution-evidence collection. Outcome pricing (O3) requires
an attribution pipeline that collects customer-visible evidence for each
billable event at definition-version granularity; capturing that
evidence at the point of the outcome event rather than reconstructing it
at invoice time is what makes the model auditable.

Meter-entitlement reconciliation and policy testing. Hybrid designs
compose a packaging anchor with an allowance mechanism and a distinct
billable unit; automated reconciliation between metered consumption and
entitlement state, together with policy tests that simulate a pricing
change against recorded usage before it takes effect, addresses the
layering that the taxonomy and the audit both identify.

The required interfaces are conventional: a rating function maps usage
events and a rate-card version to charges; an entitlement interface
answers access questions against current allowance state; an attribution
registry stores outcome events with their definition versions. None of
this requires provider-specific machinery.

\hypertarget{discussion}{%
\section{Discussion}\label{discussion}}

Read together, the findings and implications describe a two-way
interplay between software business and automation: each monetization
choice documented above creates configuration, ledger, and attribution
obligations for software systems, and the observed repricing velocity
makes discharging those obligations manually impractical.

Where credit denominations concentrate. (Interpretation.)
Credit-denominated subscription quotas are the largest AI-native share
(O1), prepaid drawdown appears at 2 of 29 observable AI-native rows
(Section 5.1), and pure metered usage remains the largest single share
in the historical panel. I read the concentration as a risk-transfer
pattern in high-variance usage domains (media generation, agent runs,
API calls): a credit-denominated quota decouples the published tier
price from shifts in underlying model cost - the provider rebalances
behind the credit rather than repricing the page - and prepayment, where
it exists, protects provider margin and cash flow. When underlying
inference costs fall while token consumption rises [15, 16], a fixed
credit quota leaves the provider free to keep or pass through any margin
improvement, since published allotments do not adjust automatically with
cost; this is an interpretation of the decoupling mechanism, not a
measured finding. The dataset captures pricing mechanics, not demand
levels, so this is an association, not a demand finding. Credits can
also provide a visible budget boundary: a prepaid balance establishes a
known amount before consumption, capping bill-overshoot risk, and a
committed quota is consumed visibly against a known limit. Where usage
is uniform or predictable, seats and flat tiers persist (O4).

Product-management reading. (Interpretation.) The findings map to three
recurring product decisions. When usage is uniform or predictable, seats
with layered meters remain sufficient (O4). When usage variance is high,
a credit-denominated quota gives the customer a visible budget boundary
and gives the provider room to rebalance unit economics behind the
published tier price. Outcome pricing becomes available only where
outcomes are observable, attributable, and contractible enough to serve
as a contractual billing event (O3) - which in this sample is customer
service and little else.

Risks and open questions. Credit expiration creates accounting-treatment
and consumer-protection considerations, particularly where unused
balances are forfeited. Outcome pricing transfers measurement trust to
the provider's definition of success. Whether credits survive margin
normalization in inference is an open question the 2027 revision of this
dataset can answer.

Compositional pricing. (Interpretation.) The independent-coder boundary
cases expose a deeper property of the observed landscape: AI pricing is
increasingly compositional rather than categorical. A packaging anchor,
an entitlement or quota, a consumption meter, a settlement mechanism,
and outcome attribution can coexist as separate layers within a single
offering, and the dual-code convention records this layering explicitly.
This is the conceptual bridge between the observed constructs and the
billing-platform implications of Section 5: a platform that models
pricing as a composition of these dimensions, rather than a single
mutually exclusive type, directly supports the hybrid designs observed
here.

\hypertarget{research-projections}{%
\section{Research projections}\label{research-projections}}

The findings above end at the September 2026 freeze. This section is my
own forward-looking interpretation, grounded in that evidence; none of
it is a measured result. Based on the observed patterns, I formulate
four falsifiable projections for future observation. Each expectation
states its evidence base, a time horizon, and the observation that would
falsify it.

F1. Hybrid structures persist. I expect most established software
companies adding AI to keep an existing anchor (subscription or
platform) and meter AI consumption on top of it, rather than moving to
pure consumption pricing. Evidence: subscription with usage quota is the
dominant observable AI-native model (O1), and established products
absorb AI cost by layering meters on seats (O4); the 2023--2026 timeline
shows layering, not migration (Section 4). Horizon: 24 months. Falsified
if more than two of the ten established-SaaS cohort companies (Section
4) drop the anchor-plus-meter structure for AI offerings in favor of
pure consumption.

F2. Credits spread as the customer-facing budget mechanism. I expect
credit denominations to keep expanding beyond AI-native sellers into
established companies' AI offerings, because a balance decouples the
published price list from shifting model costs and gives buyers a
committed budget. Evidence: credit-based entries recur across the
2023--2026 timeline (Adobe 2023, HubSpot 2024, Salesforce 2025, Notion
2026). Horizon: 24 months. Falsified if two or more of those four
abandon credit denominations for direct dollar metering.

F3. Outcome pricing grows only where outcomes are contractible. I expect
outcome-priced AI offers to concentrate in domains with an observable,
attributable, auditable event - customer-service resolution today,
possibly code-review or ticket completion next. Evidence: outcome
pricing appears only in customer service in this sample (O3). Horizon:
24--36 months. Falsified if three or more companies publish rate-carded
outcome pricing in any single domain without such an event.

F4. Seat pricing remains an important packaging anchor. Even where AI is
the headline feature, I expect per-seat pricing to remain an important
packaging anchor for many established SaaS products - a stronger claim
than F1, which allows non-seat anchors. Evidence: across the ten-company
timeline cohort (a narrative cohort assembled from dated repricing
events rather than an ex-ante sample; Section 4 and Appendix A), only
one established company (Notion) has publicly moved an AI offering from
a flat add-on to credits (Section 4). Horizon: 24 months. Falsified if
more than two of the ten established-SaaS cohort companies publicly
retire seat pricing for their AI offerings.

Design priority for billing platforms. The design weight rests on
cross-sectional evidence, not on a trend. In the September 2026
cross-section, subscription with usage quota is the largest AI-native
share (O1), robust across all 512 boundary-code assignments and under
company-level collapsing (Section 3). In the historical panel (Fig. 1),
it is present at every checkpoint - but the panel is a set of repeated
cross-sections whose observable composition changes between checkpoints,
so its shares do not measure movement: within the seven companies
observable at all four checkpoints, construct codes are stable across
three years except one transition (Gamma, seat-flat to subscription with
usage quota, between mid-2023 and mid-2024), which moves into, not away
from, the quota model. Outcome-based pricing appears in the panel only
from 2025 (10--11\%); its entries are Intercom Fin (earliest archived
outcome-coded capture June 2025), Salesforce Agentforce (April 2025),
and Sierra (September 2026, its earlier archived pages not publicly
observable); the dated public launches - Fin in March 2023, Zendesk's
announcement in August 2024 - are in the timeline (Section 4), and
Zendesk AI is not a panel member, so the 2025 appearance reflects
archive coverage, not the absence of outcome pricing earlier. These
panel shares are not directly comparable to the AI-native cross-section
shares in Section 3: the populations and denominators differ. The direct
implication for billing-platform design: represent subscription with
usage quota as a first-class model, with entitlement gating, recurring
quota resets, overage paths, and credit-denominated quotas in the
platform core rather than in add-on layers. The broader requirement is
architectural flexibility: pure metered usage is the largest single
share in the panel at every checkpoint, and prepaid credit drawdown,
seat-plus-usage hybrids, and outcome-based charging persist alongside,
so a platform must compose quota subscriptions with pure metering,
credit wallets, and outcome meters rather than optimizing for one model.
The independent-coder audit sharpens the same point from another
direction: the three audited boundary products (Notion, Vercel,
Salesforce Agentforce) each combine a packaging anchor, an allowance or
payment mechanism, and a distinct economic billable unit, so billing
platforms should model these as composable dimensions rather than one
exclusive type.

These expectations are bounded by the same limits as the rest of the
paper: public artifacts only, English-language pages, and
announcement-based timeline evidence.

\hypertarget{threats-to-validity}{%
\section{Threats to validity}\label{threats-to-validity}}

Public pricing pages are the marketing surface, not the contract;
enterprise agreements diverge from list pricing in ways this method
cannot see. Construct validity follows from this: the primitives
code what a provider publishes about how it charges, which is not
necessarily how contracts are written or invoiced, so the taxonomy
describes public pricing designs rather than billing practice. Pages change without notice; capture dates and Wayback
snapshots mitigate this, but snapshot coverage is uneven. Coding was
single-coder at capture, by the same author who designed the taxonomy;
the blinded independent re-coding of a 20\% subset (Section 2.4)
overturned no codes, but it covered only 15 of 75 rows, so judgment
variance in unaudited rows cannot be excluded; the dual-code convention
and the sensitivity band bound the rows already known to sit on
boundaries, not unknown ones. The sample skews to English-language
companies with US-visible pricing. The AI-native core is itself selected
on 2026 prominence (the full Forbes AI 50, 2026), so its cross-sectional
shares describe companies notable in 2026 rather than a defined
population of AI companies; the historical panel's survivorship
disclosure (Appendix A) applies to this core as well. The cloud-platform
cohort is enumerated by a documented rule rather than drawn from an
authoritative population; completeness of that enumeration cannot be
independently demonstrated (Appendix A). Author positionality is
disclosed in Section 2.5 and bounded by the public-sources rule. The
2026 portion of the study window is observed only through the September
16--17 capture date and is therefore right-censored: 2026 event counts
understate a full year and are not directly comparable to complete
years; timeline completeness additionally depends on the archive
reconstruction noted in Section 4.

Selective missingness. The AI-native prevalence figures are
complete-case shares, and the missingness is plausibly selective rather
than random: within the coded public purchase paths, rows with no public
pricing more often show a demo or contact-sales call to action than
observable rows (Section 3, O6), and withholding pricing is itself a
go-to-market choice. The direction of any resulting bias is unknown:
sales-led products may run more negotiated, customized structures (which
would make self-serve constructs such as subscription with usage quota
appear more common than they are among all AI-native companies) or
simply unpublished versions of the same constructs (which would leave
the shares representative). The study does not adjust for missingness;
it reports what is observable and labels it as such.

Construct validity. A first construct-validity limitation is the
distinction between public list pricing and realized customer pricing:
enterprise customers may receive negotiated discounts, committed-use
arrangements, volume pricing, or contractual terms not visible on public
pricing pages, so the study measures publicly observable pricing design
rather than realized transaction prices. The pages are also marketing
surfaces: they present the public offer as the company chooses to
represent it, and the study codes that representation, not the
implemented billing system. A second is primary-primitive assignment.
The taxonomy assigns each hybrid product a primary primitive; products
combining a packaging anchor, an allowance mechanism, and a distinct
billable unit are compressed into one category for prevalence analysis.
The dual-code convention records the layering, and the independent-coder
audit quantifies residual disagreement (Section 2.4), but primary-code
prevalence figures may simplify some hybrid designs.

Appearance of customer-knowledge exploitation. Because I work at a
billing platform, many sampled companies may be its customers. This
creates an important potential validity threat. Mitigations are
structural: the sample is fixed by external criteria (a published
ranking and documented rules), not chosen around any customer set; every
claim traces to a public artifact with URL and capture date; no
payment-provider relationships are identified or used anywhere in the
dataset or text; and the completed second-coder audit (Section 2.4)
overturned no codes.

\hypertarget{related-work}{%
\section{Related work}\label{related-work}}

Closest prior work. Bain \& Company [11] analyzes publicly available
pricing across $\sim$200 B2B SaaS (June 2026) and reports
effort/output/outcome meter shares ($\sim$35/55/10\%) and the
capacity-vs-direct-usage split ($\sim$4 in 5 capacity). It is the closest
existing analysis: multi-company, public-pricing-based, current. It
differs from this study in several respects relevant here: it publishes
aggregates without a per-company dataset or sources; it mixes general
B2B SaaS rather than defining AI-native and cloud cohorts; its taxonomy
stops at meter labels (effort/output/outcome) without coding credit
mechanics (prepaid drawdown, expirations, allotments, provisioned
throughput units (PTUs), refreshing allowances); and it draws
pricing-strategy conclusions, not software-engineering implications for
billing platforms. The prevalence numbers here are broadly consistent in
direction with Bain's - consumption-metered models dominate both
analyses, and outcome pricing is a small share in each - but the
taxonomies do not map one-to-one (Bain's effort/output/outcome meter
labels cut across this paper's primitives), so the comparison is a
directional cross-check rather than a numeric reconciliation. Open
Future Forum [12] runs a Founder AI Pricing Index from 148 founder
survey responses (usage 43\%, outcome 24\%, flat 24\%, seat 20\%,
any-mention) - self-reported and startup-skewed, but a useful
triangulation point. Metronome [13] surveys 100 SaaS companies on
UBP adoption (provider-published, survey-based). OpenView [14] is
the early adoption survey. Zhu [15] builds a reproducible public
price index over 3,208 models and 86 providers - rigorous on inference
price levels, but it studies model prices, not pricing-model design.
Microsoft Research [16] shows reasoning models raise token
consumption faster than token prices fall - adjacent cost evidence.
Fradkin [17] gives the economics framing. Bergemann, Bonatti and
Smolin [27] derive optimal pricing menus for LLM services under
variable token costs and heterogeneous users, showing that two-part
tariffs with higher markups for intensive users are optimal - a
theoretical rationale for the tiered access-plus-usage structures
observed here (Section 3). Wu, Jin and Liu [28] analyze bundling
jointly on variety and usage, showing that offering pay-per-use and
subscription side by side segments high- and low-value customers -
independent theoretical support for separating packaging from payment
mechanism in the compositional model (Section 3).

Pricing as a software-engineering artifact. The Pricing-driven DevOps
line (ISA group): García-Fernández et al. [18] map machine-readable
pricings to constraint problems over 150 pricings; [19] extracts
pricing structure from SaaS websites with LLMs; Pricing4SaaS [20]
maintains the pricing dataset these build on. This line treats pricing
as an engineering artifact to automate, not as a market phenomenon to
measure - complementary, and a source of formalism for the coding schema
used here. In the software-engineering-in-practice literature: [21]
addresses correctness of pricing-agreement enforcement; [22]
establishes the public-artifact-mining method; [23] is
ecosystem-scale empirical precedent.

Two further bodies of work are adjacent. Laatikainen, Ojala and
Mazhelis [29] propose a seven-dimension cloud pricing framework, based
on a literature review and on the pricing models of 54 cloud providers.
Saltan and Smolander [30] review 387 academic and grey-literature items
in a multivocal review of SaaS pricing and offer a taxonomy of pricing
concepts and affecting factors. Both predate the 2023--2026 period studied here. They are the nearest
SaaS and cloud pricing taxonomies, and the primitives in this paper are
intended as an AI-era complement to them, not a replacement.

Novelty statement. The research question - how AI companies price - is
not untouched: Bain's $\sim$200-company analysis and the YC index cover
prevalence. To the best of my knowledge, prior work does not combine
these four elements in a single source-traceable study: (1) a source-traceable,
row-level dataset prepared for release with the paper, subject to the
stated approval process, with dated primary sources and preserved
non-observable cases across a defined AI-native cohort (full Forbes AI
50 2026) and a rule-defined cloud-platform cohort; (2) a mechanics-level
coding (credit drawdown, expiry, allotments, refreshing allowances,
PTUs, outcome meters) recorded per row rather than meter labels, with
mechanics illustrated in Section 3 (O5) but no mechanics prevalence
reported; (3) a dated timeline of
public repricing events; (4) software-engineering implications for
billing-platform design.

\section*{Statements and Declarations}

\textbf{Funding.} The author received no specific funding for this work.

\textbf{Competing interests.} The author is employed by Stripe, a
billing and payments platform. This research was conducted independently
in a personal capacity; the views expressed are the author's own and do
not represent Stripe. The study uses only public artifacts, excludes the
author's employer from the sample, and identifies no company's payment
or customer relationships. The author declares no other competing
interests.

\textbf{Ethics approval.} Not applicable. The study analyzes publicly
available documents and pages only; it involves no human participants,
no personal data, and no non-public information.

\textbf{Data availability.} The replication dataset (coded company rows,
coding schema, codebook, timeline events, and Wayback coverage) will be
published with the paper after written Stripe approval; every coded row
carries its source URL and capture date.

\textbf{Author contribution.} The author designed the study, collected
and analyzed the data, and wrote the manuscript.

\section*{A Dataset construction and reproducibility detail}

\hypertarget{a.1-cohort-inclusion-exclusion-and-stopping-rules}{%
\subsection*{A.1 Cohort inclusion, exclusion, and stopping
rules}\label{a.1-cohort-inclusion-exclusion-and-stopping-rules}}

Cohort 1 uses the Forbes AI 50 (2026) list, chosen as the most
transparent current independent ranking (open nominations, no fees,
documented quantitative screen, conflict-recused expert judges), coded
as-is. Cohort 2 is admitted by a documented rule - a cloud or data
platform with a managed AI service and a public AI-specific pricing page
(e.g., AWS Bedrock, Google Vertex AI, Azure AI, Snowflake AI credits,
Cloudflare Workers AI) - because no independent ranking covers this
layer. Cohort 3 admits companies with a distinctive pricing design
(outcome pricing, consumptive seat add-ons, credit-quota subscriptions),
each with a written inclusion reason recorded before analysis.
Exclusions: companies without any public AI pricing artifact, ordinary
SaaS without a distinct AI SKU, and my employer. Stopping rules: Cohort
1 closes when all 50 are coded; Cohort 2 closes when every platform
identified under the rule is coded; completeness of the enumeration
cannot be independently demonstrated and is noted as a limitation
(Section 8); Cohort 3 freezes at 10 before analysis begins. After
reconciliation (ElevenLabs moved to Cohort 1 under the code-as-is rule),
Cohort 3 comprises nine rows: Adobe Firefly, DeepSeek, GitHub Copilot,
Intercom Fin, Microsoft 365 Copilot, OpenRouter, Replicate, Salesforce
Agentforce, and Zendesk AI. Earlier usage-priced APIs (GPT-3 API pricing
from 2021, per-call cloud vision and language APIs) are precursors to
the study window rather than part of the repricing shift mapped here.
Cohort sizes, stopping-rule outcomes, and per-company inclusion reasons
are in the replication dataset (Companies tab).

\hypertarget{a.2-coding-unit-and-boundary-cases}{%
\subsection*{A.2 Coding unit and boundary
cases}\label{a.2-coding-unit-and-boundary-cases}}

One row per product line: the anchor SKU is coded, and secondary models
are recorded in notes. Two companies operate two genuinely distinct
public pricing surfaces - a per-token API and a consumer subscription
app (OpenAI, Anthropic) - and each surface is coded as its own
product-line row, because a buyer encounters them as separate products
with separate pricing; this is a judgment call, disclosed here, and the
company-level sensitivity in Section 3 bounds its effect. Boundary-case
rows carry a primary and an alternate defensible code, both recorded
(Codebook tab, section 4) so external re-coding can test them.

\hypertarget{a.3-snapshot-freeze-and-update-policy}{%
\subsection*{A.3 Snapshot freeze and update
policy}\label{a.3-snapshot-freeze-and-update-policy}}

The September 16--17, 2026 snapshot is frozen. Later changes are either
Wayback reconstructions within that window or targeted corrections, each
logged in the dataset's Targeted corrections tab.

\hypertarget{a.4-archive-reconstruction-method}{%
\subsection*{A.4 Archive reconstruction
method}\label{a.4-archive-reconstruction-method}}

For the nine anchor companies, the Internet Archive Wayback Machine was
probed (CDX API, digest-collapsed, HTTP 200 captures) for the pricing
evidence URLs. Earliest archived snapshots range from October 2020
(Synthesia pricing page) to September 2025 (Claude API billing support
article). These dates mark the earliest archived observation of each
page, not the date of the pricing change. Per-company snapshot dates and
counts are in the Wayback coverage tab of the replication dataset.
Archive probing failed (rate-limited responses) for 30 of 75 company
URLs during the September 18, 2026 sweep; those entries are recorded as
probe failures, not as absence of captures. The primary pricing-page
dataset was frozen on September 16, 2026; timeline evidence collection
and verification continued through September 17, and this September 18
archive-probing sweep postdates the freeze. It reads archived captures
within the frozen window under the update policy above; it adds archive
evidence only and changes neither the primary sample nor any post-freeze
pricing state.

\hypertarget{a.5-timeline-cohort-and-historical-panel-membership}{%
\subsection*{A.5 Timeline cohort and historical panel
membership}\label{a.5-timeline-cohort-and-historical-panel-membership}}

Two groupings beyond Cohorts 1-3 support Sections 4 and 7 and are
documented here. The established-SaaS timeline cohort (Section 4)
comprises ten companies with a dated official AI monetization or
repricing announcement in 2023--2026 on an established pre-AI SaaS
product: Notion, Duolingo, Intercom, Microsoft, Adobe, Zendesk, Canva,
HubSpot, Salesforce, and GitHub. Companies whose AI announcements carry
no verified pricing change are excluded; product messaging alone is not
treated as repricing. Candidate events were identified from official changelogs, news coverage,
and company pricing pages; each retained event rests on a dated official
announcement. The cohort is assembled from the timeline rather
than by an ex-ante sampling rule; it supports the descriptive layering
observation (O4) and projections F1 and F4, not prevalence inference.
The historical panel (Fig. 1) comprises 39 companies from the 75-row
dataset for which archived pricing-page captures could be retrieved and
coded at the checkpoints; membership, snapshot dates, and per-page gaps
are recorded company-level in the replication dataset (Historical panel
and Wayback coverage tabs). It is a convenience panel skewed toward
companies notable in 2026, with explicit survivorship bias; its shares
describe the coded archived pages and do not estimate any defined
population. Only seven companies (Azure AI, Baseten, Cohere, Clay,
Gamma, Snowflake, and Synthesia) are observable at all four checkpoints;
within this balanced subset, construct codes are stable across the
checkpoints except one transition (Gamma, seat-flat to subscription with
usage quota, between mid-2023 and mid-2024).

\section*{References}

[1] Anthropic: Claude pricing.
\url{https://claude.com/pricing} (2026).
Accessed 16 September 2026

[2] Anthropic Support: How do I pay for my Claude API usage?
\url{https://support.claude.com/en/articles/8977456-how-do-i-pay-for-my-claude-api-usage}
(2026). Accessed 16 September 2026

[3] Cursor: Updates to Ultra and Pro.
https://cursor.com/blog/new-tier (2025). Accessed 16 September 2026

[4] Cursor: Clarifying our pricing.
https://cursor.com/blog/june-2025-pricing (2025). Accessed 16 September
2026

[5] ElevenLabs: Pricing.
\url{https://elevenlabs.io/pricing}
(2026). Accessed 16 September 2026

[6] GitHub Changelog: Update to GitHub Copilot consumptive billing
experience.
\url{https://github.blog/changelog/2025-06-18-update-to-github-copilot-consumptive-billing-experience/}
(2025). Accessed 16 September 2026

[7] Intercom: Fin pricing.
\url{https://fin.ai/pricing} (2026). Accessed
16 September 2026

[8] OpenAI Help Center: Setting up and managing prepaid API billing.
\url{https://help.openai.com/en/articles/8264644-setting-up-and-managing-prepaid-api-billing}
(2026). Accessed 16 September 2026

[9] Salesforce: Agentforce pricing.
\url{https://www.salesforce.com/agentforce/pricing/}
(2026). Accessed 16 September 2026

[10] Snowflake: Service Consumption Table.
\url{https://www.snowflake.com/legal-files/CreditConsumptionTable.pdf}
(2026). Accessed 16 September 2026

[11] Bain \& Company: AI Pricing: A Reality Check on Effort, Usage,
and Outcomes.
https://www.bain.com/insights/ai-pricing-a-reality-check-on-effort-usage-and-outcomes/
(2026). Accessed 16 September 2026

[12] Open Future Forum: YC Founder AI Report, September 2026: How AI
Startups Price, by Batch and by Buyer.
https://openfutureforum.com/research/yc-founder-ai-report-september-2026
(2026). Accessed 16 September 2026

[13] Metronome: State of Usage-Based Pricing 2025.
https://metronome.com/state-of-usage-based-pricing-2025 (2025). Accessed
16 September 2026

[14] OpenView Partners: Usage-Based Pricing: The Next Evolution in
Software Pricing. https://openviewpartners.com/usage-based-pricing/
(2022). Accessed 16 September 2026

[15] Zhu, L.Y.: The Price of Intelligence: A Quality-Adjusted Price
Index for AI Services. arXiv:2608.29843.
https://arxiv.org/abs/2608.29843 (2026). Accessed 16 September 2026

[16] Chen, L., Zhang, C., He, Y., Stoica, I., Zaharia, M.A., Zou,
J.: The Price Reversal Phenomenon: When Cheaper Reasoning Models End Up
Costing More. Microsoft Research.
https://www.microsoft.com/en-us/research/publication/the-price-reversal-phenomenon-when-cheaper-reasoning-models-end-up-costing-more/
(2026). Accessed 16 September 2026

[17] Fradkin, A.: The Emerging Market for Intelligence: How Firms
Buy and Sell AI. JEP preprint.
https://andreyfradkin.com/assets/jep\_llm\_preprint.pdf (2026). Accessed
16 September 2026

[18] García-Fernández, A., Parejo, J.A., Trinidad, P., Ruiz-Cortés,
A.: Automated Analysis of Pricings in SaaS-based Information Systems.
CAiSE 2025. arXiv:2503.21444. https://arxiv.org/abs/2503.21444 (2025).
Accessed 16 September 2026

[19] Cavero, F.J., Alonso, J.C., Ruiz-Cortés, A.: From Static to
Intelligent: Evolving SaaS Pricing with LLMs. arXiv:2507.12104.
https://arxiv.org/abs/2507.12104 (2025). Accessed 16 September 2026

[20] ISA Group: Pricing4SaaS dataset.
https://github.com/isa-group/Pricing4SaaS (2025). Accessed 16 September
2026

[21] Kolesar, J.C., Lepoint, T., Schäf, M., Visser, W.: Safe
Validation of Pricing Agreements. ICSE 2025 SEIP, pp. 101-111.
https://conf.researchr.org/details/icse-2025/icse-2025-software-engineering-in-practice/1/Safe-Validation-of-Pricing-Agreements
(2025). Accessed 16 September 2026

[22] Li, H., Bezemer, C.-P., Hassan, A.E.: Software Engineering and
Foundation Models: Insights from Industry Blogs Using a Jury of
Foundation Models. ICSE 2025 SEIP.
https://conf.researchr.org/details/icse-2025/icse-2025-software-engineering-in-practice/54/Software-Engineering-and-Foundation-Models-Insights-from-Industry-Blogs-Using-a-Jury
(2025). Accessed 16 September 2026

[23] Cui, X., Wu, J., Wu, Y., Wang, X., Luo, T., Qu, S., Ling, X.,
Yang, M.: An Empirical Study of License Conflict in Free and Open Source
Software. ICSE 2023 SEIP, pp. 495-505.
https://conf.researchr.org/details/icse-2023/icse-2023-SEIP/42/An-Empirical-Study-of-License-Conflict-in-Free-and-Open-Source-Software
(2023). Accessed 16 September 2026

[24] Anthropic (@AnthropicAI on X): Weekly rate limits for Claude
subscriptions. \url{https://x.com/AnthropicAI/status/1949898502688903593}
(2025).\newline Archived at
\url{http://web.archive.org/web/20260812131954/https://x.com/AnthropicAI/status/1949898502688903593}.
Accessed 16 September 2026

[25] Forbes: AI 50 (2026). https://www.forbes.com/lists/ai50/
(2026).\newline Archived at
\url{http://web.archive.org/web/20260917185914/https://www.forbes.com/lists/ai50/}.
Accessed 16 September 2026

[26] OpenAI Developer Community: OAI API switching to pre-paid
billing.
https://community.openai.com/t/oai-api-switching-to-pre-paid-billing/659398
(2024). Accessed 16 September 2026

[27] Bergemann, D., Bonatti, A., Smolin, A.: The Economics of Large
Language Models: Token Allocation, Fine-Tuning, and Optimal Pricing.
Cowles Foundation Discussion Paper No. 2425, Yale University.
https://cowles.yale.edu/sites/default/files/2025-02/d2425.pdf (2025).
Accessed 19 September 2026

[28] Wu, C., Jin, C., Liu, Q.: Bundling Variety, Usage, or Both? A
Multi-Service Analysis of Pay-Per-Use and Subscription Pricing.
Production and Operations Management 33(10), pp. 1979-1996.
https://doi.org/10.1177/10591478241265480 (2024). Accessed 19 September
2026

[29] Laatikainen, G., Ojala, A., Mazhelis, O.: Cloud Services Pricing
Models. In: Software Business. From Physical Products to Software
Services and Solutions (ICSOB 2013), LNBIP 150, pp. 117-129. Springer
(2013). \url{https://doi.org/10.1007/978-3-642-39336-5_12}

[30] Saltan, A., Smolander, K.: Bridging the state-of-the-art and the
state-of-the-practice of SaaS pricing: A multivocal literature review.
Information and Software Technology 133, 106510 (2021).
\url{https://doi.org/10.1016/j.infsof.2021.106510}

\end{document}